\documentstyle[12pt]{article}
\begin{document}

\setlength{\baselineskip}{24pt}
\newcommand{\ec}{\mbox{\( \rm e^{\rm c}\)}}
\newcommand{\e}{\mbox{e}{}}

\newcommand{\bea}{\begin{eqnarray}}
\newcommand{\eea}{\end{eqnarray}}
\newcommand{\be}{\begin{equation}}
\newcommand{\ee}{\end{equation}}

\newcommand{\nue}{\mbox{\(\nu_{\rm e}\)}}
\newcommand{\numu}{\mbox{\(\nu_\mu\)}}
\newcommand{\nutau}{\mbox{\(\nu_\tau\)}}
\newcommand{\munu}{\mbox{\(\mu_\nu\)}}
\newcommand{\mnu}{\mbox{\(m_\nu\)}}
\newcommand{\mnutau}{\mbox{\(m_{\nu_\tau}\)}}
\newcommand{\mnua}{\mbox{\(m_{\nu_a}\)}}
\newcommand{\mea}{\mbox{\(m_{{\rm e}_a}\)}}

\newcommand{\Gf}{\mbox{\(G_{\rm F}\)}}
\newcommand{\muB}{\mbox{\(\mu_{\rm B}\)}}
\newcommand{\mPl}{\mbox{\(m_{\rm Pl}\)}}
\newcommand{\Mpc}{{\rm Mpc}}
\newcommand{\MeV}{{\rm MeV}}

\newcommand{\mW}{\mbox{\(m_{\rm W}\)}}
\newcommand{\me}{\mbox{\(m_{\rm e}\)}}
\newcommand{\Wangle}{\mbox{\(\theta_{\rm W}\)}}
\newcommand{\pF}{\mbox{\(p_{\rm F}\)}}
\newcommand{\dg}{\mbox{\( |\:\rangle\)}}
\title{
Flavor-Changing Magnetic Dipole Moment
and Oscillation of a Neutrino in a Degenerate Electron Plasma
}
\author{\vspace*{0.5em}\\
Hisashi Kikuchi \\
\vspace*{0.5em}\\
{\normalsize \sl Ohu University}\\
{\normalsize \sl Koriyama, 963 Japan}
}
\date{OHU-PH-9508}

\vfil

\maketitle

\vfil

\begin{abstract}
The standard model prediction for a magnetic dipole moment of a neutrino is
proportional to the neutrino mass and extremely small.
It also generates a flavor-changing process, but the GIM mechanism
reduces the corresponding amplitude.
These properties of  a neutrino magnetic moment change drastically
in a degenerate electron  plasma.
We have shown that an electron-hole excitation  gives a contribution
proportional to the electrons' Fermi momentum.
Since this effect is absent in \(\mu\) and \(\tau\) sector,
the GIM cancellation  does not work.
The magnetic moment induces a neutrino oscillation if a strong enough
magnetic field exists in the plasma.
The required magnitude of the field strength that affects the \nue\ burst
from a supernova is estimated to be the order of \(10^8 \) Gauss.
\end{abstract}
\vfil

\newpage

A magnetic dipole moment of a neutrino induces interesting phenomena,
such as a spin rotation when its travelling in a static magnetic field
\cite{fuj,vol}
or a transition radiation when passing an interface between two different
media \cite{sak}.
It may also affect  the stellar cooling by the decay of plasmons \cite{sut},
which is known as a dominant cooling process in a dense star
\cite{ada,bea}.
The standard model predicts a nonzero value for it through the processes
depicted in Fig.~1,  but the magnitude is far below the one that current
experiments can detect.
To leading order in \(m_l^2/m_W^2\), the result is independent of the
mass \(m_l\) of the charged leptons in the internal lines;
it is instead proportional to the neutrino mass \mnu\ \cite{lee1},
\be \munu = {3 e \Gf \mnu \over 8 \sqrt 2 \pi^2} = 3 \times 10^{-19}
\left( {\mnu \over 1 \rm eV} \right) \muB, \ee
where \muB\ is the Bohr magneton.
There are many orders of magnitude between this prediction and
the present empirical limits,
\(\munu < 10^{-6} - 10^{-11} \muB \) \cite{fuk}.

The leptonic charged current coupling to the W boson generates a generation
mixing for massive neutrinos as is the case for the quark sector.
The processes in Fig.~1 can then  generate, so to say,
a flavor-changing electromagnetic current.
The independence of the leading contribution  from \(m_l\),
however, subjects the corresponding amplitude to the GIM suppression
 \cite{lee1}:
The sum of the leading contributions from all three generations
 cancels each
other because of the unitarity of the leptonic CKM
mixing matrix.
Thus the amplitude, such as the one for the decay  \(\numu \rightarrow
\nue \, \gamma\), gets a nonzero contribution
from the  next-to-leading effect and
is further suppressed by a factor \((m_l^2/m_W^2)\) than one estimates
naively with \munu.

The purpose of this letter is to show that these properties of a neutrino
magnetic moment change drastically in  a degenerate electron plasma.
In the gas of high density and relatively low temperature, i.e., where
the Fermi momentum \pF\ is much larger than the temperature \(T\),
most of the electrons degenerates into the Fermi sphere.
Electromagnetic potential \(A_\mu\) excites one of the electrons out
of the Fermi sphere and leaves a hole in it.  Subsequently the excited
electron  comes back into the sphere
emitting a pair of neutrinos by exchanging a W boson with the hole.
This polarization of a electron-hole pair turns out to add
 a contribution of the order of
\(e \Gf\, \pF\) to the magnetic moment (See Eqs.~(\ref{mu}) and (\ref{fm})
below).
Since this effect is intrinsic to degenerate electrons
 and absent in the \(\mu\)
 and \(\tau\) sectors, the  GIM  mechanism no longer washes out this
contribution.

The Fermi momentum of the electron gas at the core of massive stars
becomes as large as or larger than the electron mass \me\
at the later stage of their evolution.
This gives a possibility that the resulting magnetic
moment becomes so large
that  some observations can detect it.
We will consider a neutrino oscillation induced by the flavor-changing
magnetic moment,  which may  take place in the stars.
Our calculation uses a zero temperature approximation,  and the result
should be applied  to the cases of \(\pF \gg T\).
In the following,  we will first describe the result for the magnetic
moment briefly and then consider the induced neutrino oscillation.

We assume the masses  that  the neutrinos,
\nue, \numu,  and \nutau, get at the electroweak
symmetry breaking are of Majorana type.%
\footnote{
For the case of Dirac neutrinos,  one readily gets the corrections
by  simply replacing  \(a_{\vec p, s}^\dagger(\nu) \)
in (\ref{ex}) with \(b_{\vec p, s}^\dagger(\nu) \),  the creation operator for
the anti-neutrino state.
It does not affect our discussion in this letter.}
The corresponding fields,  which we collectively denote by \(\nu(x)\)
in the left-handed two-component notation,
have an expansion
\be \nu(x) = {1\over \sqrt V} \sum_{\vec p, s} \left[ e^{-i E(\vec p\,) x^0 +i
\vec p\cdot \vec x} \,u(\vec p, s)
\,a_{\vec p, s} (\nu) + e^{iE(\vec p\,) x^0 -i \vec p\cdot \vec x}\,
v(\vec p, s) \, a^\dagger_{\vec p, s} (\nu) \right]
\label{ex} \ee
in the annihilation, \(a_{\vec p,s}(\nu)\),  and the creation, \(a_{\vec p,
s}^\dagger(\nu) \), operators of the state with the momentum
\(\vec p\) and the helicity \(s\);
the spinors \(u(\vec p, s)\) and \(v(\vec p, s)\) are defined by
\bea u(\vec p, s) &= &\sqrt{ E(\vec p\,) - s |\vec p\,| \over 2 E(\vec p\,) }
\chi(\hat p, s) \\
v(\vec p, s) &= &- \sqrt{ E(\vec p\,) + s |\vec p\,| \over 2 E(\vec p\,) }
\epsilon \chi(\hat p, s)^*,
\eea
where
\(E(\vec p\,) = \left( {\vec p\,{}^2 + \mnu^2}\right)^{1/2}\),
\(\chi(\hat p, s) \) is the normalized eigenspinor for the helicity,
\be (\hat p \cdot \vec\sigma) \, \chi(\hat p, s) = s \,\chi(\hat p, s),
\quad \chi^\dagger(\hat p, s) \chi(\hat p, t) = \delta_{st},\ee
defined with the Pauli matrices \(\vec \sigma\), and
\( \epsilon \equiv i \sigma^2 \).
The quantisation volume \(V\) in (\ref{ex}) should be taken as the size
of the plasma.
Note that for states with relativistic momentum,  \(\nu(x)\) is dominated
by annihilation operators of \( s = -1\) and creation operators of \(+1\).

The electrons and positrons are described by \( \e(x)\) and \(\ec(x)\) in the
two-component notation.
\( \e(x)\) has the charged current coupling to a W boson, while
\(\ec(x)\) does not.
They have  the  same  expansion as Eq.~(\ref{ex}) if one does proper
replacements; in \(\e(x)\) the annihilation operators stand for
electrons, \(a_{\vec p,s}({\rm e})\),  and
the creation operators for positrons, \(b_{\vec p,s}^\dagger ({\rm e})\),
while  in  \(\ec(x) \)  the annihilation operators stand for positrons,
\(b_{\vec p,s}({\rm e})\), and the creation operators for electrons, \(a_{\vec
p,s}^\dagger ({\rm e})\);
the energy in the expressions of  \(u(\vec p, s) \) and \(v(\vec p, s) \) is
understood to be calculated with  \me.

In terms of these fields,  the one-loop induced electromagnetic
vertex for the neutrinos has the form
\be i{\cal L}_{\rm eff} = \left( {-i e g^2\over 2 }\right)
V_{\nu_a \rm e} V^*_{\nu_b \rm e} \nu_a^\dagger (p_1) F^\mu(p_1,p_2)
\nu_b(p_2)
A_\mu(q), \label{L}\ee
where,  \(e\) and \(g\) are the electromagnetic
and charged current coupling constants,
\(V_{\nu_a \rm e}\) is the element of the CKM  matrix between
\(\nu_a\) (\( a = \e, \mu, \tau\)) and \e, and the fields are written in their
Fourier components
of momenta \(q\), \(p_1\), and \(p_2\), that  satisfy \(q + p_1+ p_2 = 0\).
The structure function \(F^\mu\) in Eq.~(\ref{L}) is given from
the expectation value of a T-product,
\bea F^\mu(p_1, p_2) &\equiv& \int\int dx dy e^{ -i p_1 x -i p_2 y }
\nonumber\\
&&\times
\langle\:| \, {\rm T}\,  W^+_\nu(x) \bar\sigma^\nu \e(x) \, \left[ e^\dagger(0)
\bar\sigma^\mu e(0)
- \ec^\dagger(0) \bar\sigma^\mu \ec(0) + ... \right] \,
\e^\dagger(y) \bar\sigma^\lambda W^-_\lambda (y) |\:\rangle,\label{F}
\eea
where the state
\be | \: \rangle = \prod_{|\vec p\, |<p_{\rm F},  s=\pm1}
a_{\vec p, s}^\dagger (\rm e) |0\rangle\ee
represents the degenerate electron plasma;
\(\bar\sigma ^\mu   \equiv (1, - \vec \sigma) \).
We have abbreviated the contribution from W's electromagnetic vertex in
Eq.~(\ref{F}).
The calculation is carried out by  modifying the one  for
a non-relativistic plasma \cite{fet}.
The usual Feynman rule applies if one uses the propagators that take
Pauli exclusion principle into account.
They are obtained from Eq.~(1) in Ref.~\cite{ada}.
Since we are interested in the effect proper  to a degenerate plasma,
we subtract the  vacuum contribution \(F^\mu_{\rm vc}\)
from \(F^\mu\) and concentrate on the  remaining term \(F^\mu_{\rm dg} =
F^\mu - F_{\rm vc}^\mu \)
(\(F^\mu_{\rm vc}\) is defined by \(F^\mu\) with zero \pF).

We first specify the form of \(F^\mu_{\rm dg}\) taking various conditions
into account.
Since we have assumed the plasma is isotropic and homogeneous,
the temporal component \(F^0_{\rm dg}\) is  a scalar while the spacial
components \(\vec F_{\rm dg}\) are vectors.
To leading order in \(1/\mW\), only the  contribution of
Fig.~1 (a)  remains  and the W boson propagator can be safely contracted to a
point form,  \(G_{\mu\nu} \sim  g_{\mu\nu}/\mW^2\).
This is because the loop momentum  is restricted by \pF\ in \(F^\mu_{\rm
dg}\).
Thus \(F^\mu_{\rm dg}\) depends only on  \(q\).
There are two structures, \(1\) (unit matrix) and \(\vec\sigma\cdot\vec
q\), for \(F^0_{\rm dg}\) and four structures, \(\vec q\),
\((\vec\sigma\cdot\vec
q\,)\vec q\), \(\vec \sigma\), and \(\vec\sigma\times\vec q\), for
\(\vec F_{\rm dg}\).
They are also constrained by the gauge invariance, \(q^0 F^0_{\rm dg} - \vec
q\cdot \vec F_{\rm dg} =
0\).
Finally, we recall  the study on
\(F^\mu_{\rm dg}\)  in Ref.~\cite{ada} for the plasmon decay.
Although it was done with the four Fermi interaction,
the result still applies because the W boson propagator is
safely  contracted.  The four Fermi interaction is Fierz-transformed into
the product of the neutral current of  neutrinos and the  V--A current of
electrons.
\(F^\mu_{\rm dg}\) is then  divided into two components,
one from  electrons' vector current and the other from their axial vector
current \cite{ada}.
The vector component is related by
\(F^\mu_{\rm dg} \sim \Pi^\mu_\nu\bar\sigma^\nu/2 \mW^2\)
to the polarization tensor \(\Pi_{\mu\nu}\) and, thus,
turns out to consist of  two independent structures.
For the axial component, we are left with only  one available structure
\(\vec\sigma
\times \vec q\).
We realize that \(F^\mu_{\rm dg}\)  has the form
\bea
F^0_{\rm dg} &=&  f_{\rm l} \left[ |\vec q\,|^2  +q^0  (\vec \sigma \cdot
\vec q\,) \right], \\
\vec F_{\rm dg} & =  & f_{\rm l} \, q^0 \vec q\, \left[  1
+ { q^0 \over |\vec q\,|^2  } (\vec \sigma \cdot
\vec q\,) \right] +  f_{\rm t}\, \left[ |\vec q\,|^2 \vec \sigma
- (\vec\sigma\cdot\vec q\,)\vec q\,)\right] +
f_{\rm m} \, i (\vec \sigma \times \vec q\,)
\eea
with three form factors \(f_{\rm l}\), \(f_{\rm t}\), and \(f_{\rm m}\),
which depend on the rotational invariants \(|\vec q\,|\) and \(q^0\).%
\footnote{
As far as the decay of a plasmon is concerned, \(f_{\rm l}\) and \(f_{\rm
t}\) give dominant contribution over \(f_{\rm m}\) \cite{ada} (See also
\cite{koh}).}
Applying the time reversal invariance of \dg,
we see that all of the form factors are an even function of  \(q^0\).

The flavor-changing magnetic dipole moment \(\mu_{ab}\) is related to
\(f_{\rm m}\) by
\be \mu_{ab} = {eg^2\over 2} V_{\nu_a {\rm e} } V_{\nu_b {\rm e} }^*
f_{\rm m}, \label{mu}\ee
where we have used the definition that the corresponding Lagrangian is
written as
\be {\cal L}_{\rm m} = \mu_{ab} (\nu_a^\dagger \vec \sigma \nu_b ) \cdot
\vec B, \label{Lm}\ee
with the magnetic field \(\vec B\).
Under the  CP transformation, \(f_{\rm m}\) is odd (while \(f_{\rm l}\) and
\(f_{\rm t}\) are even).
That means \(\mu_{ab}\) changes its sign if the plasma is made of  positrons,
or positrons and electrons contribute destructively to \(\mu_{ab}\) if they
coexist in a
plasma.

We evaluated \(I \equiv {\rm Tr}(\sigma^j F^i_{\rm dg})/2 \)
to extract out \(f_{\rm m}\).
Keeping only the terms that are proportional to the anti-symmetric tensor
\(\epsilon_{ijk}\),  we found
\bea I& =&  i \epsilon_{ijk} \left( {1\over  \mW^2}\right) \int {d\vec p
\over
(2\pi)^3}\nonumber\\
&& \times\left[
\left( { p^k + q^k \over E(\vec p + \vec q\,)}- { p^k \over E(\vec p\,)}
\right)
{\theta( | \vec p + \vec q\,| - \pF ) \theta(\pF - | \vec p\,| ) (E(\vec p +
\vec
q\,)- E(\vec p\,))
\over (q^0)^2- ((E(\vec p + \vec q\,)- E(\vec p\,))^2 }\right.\nonumber \\
&&- \left. \left( { p^k + q^k \over E(\vec p + \vec q\,)}
+ { p^k \over E(\vec p\,)} \right) {\theta(\pF - |\vec p + \vec q\,|)
(E(\vec p + \vec q\,)+ E(\vec p\,))
\over (q^0)^2- ((E(\vec p + \vec q\,) + E(\vec p\,))^2 }\right] , \eea
where \( E(\vec p\,) = ( \vec p\, ^2 + \me^2)^{1/2}\).
The first term comes from the electron-hole excitation;
the second represents the vacuum polarization of a electron-positron
pair that is now forbidden
if the electron has the momentum in the Fermi sphere.
We are interested in the static component of \(\mu_{ab}\);  for \( q^0 = 0\)
and \( |\vec q\, | \ll \pF\),  we found
\be f_{\rm m } = - {1\over 4\pi^2} {\pF\over \mW^2}. \label{fm}\ee
Details of the calculation will be presented elsewhere \cite{kik}.

Let us turn to a neutrino oscillation induced by \( \mu_{ab}\).
The oscillation we are considering  here is the one where a neutrino
in one of the flavors (mass eigenstates)  oscillates into another flavor
in the presence of an external magnetic field \(\vec B\).
Thus,  it is conceptionally different from the vacuum oscillation \cite{mak}
in which
a neutrino, created in an eigenstate of the weak interaction coupled
to  an electron, oscillates into another kind.
We assume the neutrino energy \(E\) is relativistic, \(\mnu/E\ll 1\),
in the Lorentz flame where the plasma has zero mean velocity.
We also assume the deviation of \(\vec B\) from
the  completely static and homogeneous configuration  is small, and
the energy and momentum transfer from \(\vec B\) to the neutrino is
negligible compared with \(E\).

In a vacuum,  where a flavor-changing process is suppressed,
a physically interesting process induced by a magnetic moment is an
oscillation between two different helicities.
This is the spin rotation discussed in Refs.~\cite{fuj,vol}.
This helicity-flipping process is, however,  suppressed by the factor
\((\mnu/E)\)
as one can immediately see from the explicit form for the coupling,
Eq.~(\ref{Lm}),  and the expansion, Eq.~(\ref{ex}).%
\footnote{The proportionality of \munu\ to \mnu\ comes from this factor
in our two component notation.}
An advantage  in a degenerate plasma is that an oscillation is possible
between two different flavors keeping the  helicity intact and thus without
the suppression of \(\mnu/ E\).

For the sake of clearness of the argument,  we specifically
consider an oscillation between \nue\ and \numu\ with \(s = -1\).
We adopt a two-state approximation; we restrict the Hilbert space with
two states, \( |\nue \rangle = a_{\vec p_1, -1 }^\dagger (\nue)\dg \)
 and \(|\numu  \rangle = a_{\vec p_2, -1 }^\dagger(\numu)\dg \),
and consider  an oscillation just between them.
Based on the assumptions mentioned above, we assume the energies are the
same.
Then the momenta (whose magnitudes are \(\sqrt{ E^2 - m_{\nu_e}^2 } \) and
\(\sqrt{ E^2 - m_{\nu_\mu}^2}\)) should be slightly different, which is taken
account for by a small momentum transfer \(\vec q\,\) from \(\vec B\).

The Hamiltonian \(H\),  a \(2\times 2\)  matrix in our approximation, get
off-diagonal matrix elements in the presence of \(\vec B\).
It reads
\be
H = \left( \begin{array}{cc} E&A\\A^*&E\end{array}\right), \ee
where
\be A = \int_V d\vec x \langle\nue|(-{\cal L}_{\rm m})|\numu\rangle \simeq
 \mu_{{\rm e} \mu}  \cos \theta |\vec B|,\quad \hat p\cdot \hat
B = \cos \theta.\ee
The new eigenstates are a mixture of \(|\nue\rangle\) and
\(|\numu\rangle\) with  equal weights and a different relative phase;
their energy difference is
\( \Delta E = 2 |A| \).
The half-oscillation length \(L^{(1/2)} = \pi/2 \Delta E\) characterizes the
oscillation; the  relativistic neutrino that is initially in
 \( |\nue\rangle\) gets
a fifty percent probability to be detected as \(|\numu\rangle\)
after a travel of this length.

We mention a few of the characteristics of \(L^{(1/2)} \),
other than the absence of the suppression of \(\mnu/E\) mentioned above.
First, it does not depend on the energy or the mass difference
of the neutrinos, while in the vacuum oscillation the oscillation length is
given by \(\pi E/ (m_{\nu_\mu}^2 - m_{\nu_{\rm e}}^2 )\).
Secondly, it inversely proportional to \(\cos\theta\);
the oscillation takes place most efficiently for the momenta nearly
parallel to \(\vec B\), while the spin rotation is most efficient
when the neutrino travels perpendicular to \(\vec B\) \cite{fuj}.

The stellar interiors are the candidates where the
magnetic-moment-induced
neutrino oscillation may have  a physical importance.
We relate \pF\ to the mass density \(\rho\) of the interior and
the electron's number fraction \(Y_{\rm e}\) per baryon,  and write the
oscillation length as
\be L^{(1/2)} = 6.1 \times 10^{17} \, \mbox{cm}\,[  Y_{\rm e}\, \rho( {\rm
g/cm^3})] ^{-1/3}\,
[|\vec B|({\rm G})]^{-1} \,[ |V_{\nu_e e} V_{\nu_\mu e}^* \cos\theta|]^{-1}
\ee
with values of \(\rho\) in units of gram per cubic centimeter and
of \(|\vec B\,| \) in Gauss.
Our zero temperature calculation applies for the case
\( \pF \gg T\), which reads
\([  Y_{\rm e}\,\rho( {\rm g/cm^3})] ^{1/3} \gg
1.7 \times 10^{-8} \, T(\rm K)\) with \(T\)  in units of Kelvin.
For the solar neutrinos, the oscillation is too slow;
even taking optimistic values,  \(\rho \sim 10^2 \, {\rm g/cm^3}\) and
\( |\vec B\,| \sim 10^3 \, {\rm G} \),  we get  \(L^{(1/2)} \sim 10^{14}\,
{\rm cm}\,
Y_{\rm e}^{-1/3} [ |V_{\nu_e e} V_{\nu_\mu e}^* \cos\theta | ]^{-1} \),
which is much larger  than the solar radius \(\sim 7\times 10^{10} {\rm
cm}\).

An interesting possibility is that the oscillation may convert \nue\
from a supernova to \numu\ or \nutau.
A massive star, whose mass is bigger than \(8 M_\odot\), has its core
consist mainly of the Fe elements at the
final stage of its evolution.
The degenerate electron plasma plays an important role to keep the core
from collapsing.
It, however, fails when the mass of the core exceeds the Chandrasekhar
mass.
The core begins to collapse and \nue\ is copiously produced by the electron
capture  by the Fe nuclei.
The collapse eventually stops when the central density reaches
to the nuclear density, and a shock wave is formed at the central
region  of the core.
The shock wave then produces the \nue\ burst by the neutronization of free
protons when it propagates outward in the core \cite{suz}.
The \nue\ flux generated by these processes
in the very early stage of the explosion
is  converted to another kind  if there penetrates
a strong enough magnetic field in the core;
the resulting flux becomes a composition of \nue\ and a substantial amount
of \numu\ or \nutau.
We estimate the required field strength for this to happen by the condition
that \(L^{(1/2)}\) is shorter than \(R_{\rm core}\), the core radius.
Using the values \cite{suz},   \(R_{\rm core} \sim 10^7 {\rm cm}\) and
\(\rho \sim 10 ^{10} {\rm g/cm^3}\),
we find
\( |\vec B\,| > 10^8 \, {\rm G} \,Y_{\rm e}^{-1/3} [ |V_{\nu_e e} V_{\nu_\mu
e}^* \cos\theta | ]^{-1} \).
This magnitude seems realizable  if one compares it with the value
\(|\vec B\,| \sim 10^{13} \, {\rm G}\) which is possible for a
supernova or a neutron star as  discussed in Ref.~\cite{fuj}.

\newpage
\begin{center}
{\bf Figure Caption}
\end{center}
Fig.~1: The Feynman diagrams for the magnetic dipole moment of a neutrino
 in the standard model.
\end{document}